# A cheap liquid additive from acetone for strong in-field $J_c$ enhancement in MgB$_2$ conductors

Dongliang Wang[1], Yanwei Ma[1, #], Zhaoshun Gao[1], Xianping Zhang[1], Lei Wang[1], S. Awaji[2], K. Watanabe[2], E. Mossang[3]

[1] Key Laboratory of Applied Superconductivity, Institute of Electrical Engineering, Chinese Academy of Sciences, P. O. Box 2703, Beijing 100190, People's Republic of China

[2] High Field Laboratory for Superconducting Materials, Institute for Materials Research, Tohoku University, Sendai 980-8577, Japan

[3] Grenoble High Magnetic Field Laboratory, C.N.R.S., 25, Avenue des Martyrs, B.P. 166, 38 042 Grenoble Cedex 9, France

**Abstract:**

We report on significant flux pinning enhancement in MgB$_2$/Fe tapes that has been easily obtained by a simple and cheap route using acetone as both an efficient ball-milling medium and liquid additive through the in situ method. Results showed that the highly reactive C released from the decomposition of the acetone substituted into B sites, accompanied by the grain refinement effect due to the acetone doping. At 4.2 K, the transport $J_c$ for the 5 wt % acetone doped tapes sintered at 700 °C reached up to $2.4 \times 10^4$ A/cm$^2$ at 10 T, which is even higher than that of the nano-C added samples heated at 900 °C.

[#] Author to whom correspondence should be addressed; E-mail: ywma@mail.iee.ac.cn

## 1. Introduction

Since the discovery of superconductivity in $MgB_2$[1], enormous effects have been directed towards improving the in-field critical current density ($J_c$) and the upper critical field ($H_{c2}$). It has been demonstrated that the $J_c$ of $MgB_2$ can be significantly enhanced by carbon-based nanoparticles doping, such as SiC,[2–4] C,[5,6] carbon nanotube,[7] and $B_4C$.[8] These nanodopants could been introduced through solid state mixing. However, for this method, it is a great challenge to achieve homogeneous distribution of a small amount of nanoadditives and the matrix materials.[9]

By far, a series of methods were employed to solve the problem of nanoparticles agglomeration, such as the wet mixing method[9-13] and the use of the liquid additives.[14–17] Zhou et al achieved the $J_c$ enhancements by "wet" mixing of a sugar solution with a boron powder and successive drying to achieve homogeneous doping.[10] The excellent high-field performance in Fe-sheathed $MgB_2$ tapes has also been achieved though using inexpensive stearic acid and stearate additives.[11,12] In comparison with the wet mixing method, the liquid doping method is a simple route which does not need the drying process. Silicon oil, as a liquid additive, has been claimed to solve the problem of nanoparticle agglomeration, and resulted in a significant enhancement of $J_c$, $H_{c2}$, and $H_{irr}$.[14] It has also been reported that acetone could be a carbon source to significantly enhance the $H_{c2}$ and $H_{irr}$ of $MgB_2$ bulks.[15] Meanwhile, homogeneous distribution of nanoscale dopant and the starting powders was also achieved just by ball milling the precursor powders immersed in acetone solution, which avoids the direct contact between the mixture powders and $O_2/H_2O$.[16] However, the effect of acetone doping on $J_c$-$H$ properties of Fe-sheathed $MgB_2$ tapes has not been reported. Therefore, we expect that the high $J_c$-$B$ performance of $MgB_2$ tapes fabricated by a simple and cheap route can be achieved by ball-milling the precursor powders with acetone. In this work, the effects of the acetone dopant on the microstructure, phase identification, and transport properties of Fe-sheathed $MgB_2$ tapes are investigated. The mechanism for the enhancement of flux pinning is also discussed.

## 2. Experiment details

$MgB_2$/Fe tapes were prepared with spherical magnesium powders (10 $\mu$m,

99.5%) and boron powders (1-2 $\mu$m, 99.99%) by the in-situ PIT process. The magnesium powder and boron powder at chemical stoichiometry were thoroughly mixed within the acetone. The amounts of acetone ($CH_3COCH_3$) added into the powders were 5, 10, and 15 wt %, respectively. In order to avoid the evaporation of the acetone, the powders were mixed using the ball-milling method. The ball-milling process was carried out for 1 h with a rotating speed of 400 rpm under air in a planetary ball mill using agate vials and balls. The mixed powders were filled into pure Fe tubes of 8 mm outside diameter and 1.5 mm wall thickness in air. The tubes were subsequently rotate swaged, drawn and rolled into tapes. Further experiment details were described elsewhere.[18] Finally, short samples of about 40 mm were heat-treated at 600–800 °C for 1 h, and then followed by cooling down to room temperature in the furnace. The argon gas was allowed to flow into the furnace during the heat-treatment process to reduce the oxidation of the samples. An undoped tape was also prepared under the same conditions for use as a reference sample.

The phase identification was performed by x-ray diffraction (XRD) using Cu K$\alpha$ radiation. For study of tapes, the Fe sheaths were mechanically removed to expose the core. Microstructural observation was carried out by scanning electron microscopy (SEM) and transmission electron microscopy (TEM). The DC magnetization measurement was performed with a superconducting quantum interference device (SQUID) magnetometer. The $T_c$ value was determined by taking the first deviation point from linearity that signifies the transition from the normal to superconducting state. Transport critical current ($I_c$) of the samples were measured at 4.2 K and its magnetic field dependence were evaluated at the High Field Laboratory for Superconducting Materials (HFLSM) in Sendai and Grenoble High Magnetic Field Laboratory in France, by using a standard four-probe technique, with a criterion of 1 $\mu$V/cm. A magnetic field up to 14 T was applied parallel to the tape surface. For each set of tape made from different types of precursor powders, the $I_c$ measurement was performed for several samples to check reproducibility.

### 3. Results and discussion

Figure 1 shows the transport $J_c$ at 4.2 K in magnetic fields up to 14 T for 5 wt %

acetone doped samples heated at different temperatures ranging from 600 to 800 °C. The $J_c$ values of the pure and nano-C-doped tapes made by ball milling without acetone are also included.[18] Surprisingly, all the 5 wt % acetone doped samples sintered at different temperatures exhibited superior $J_c$ values compared to the pure tape.[18] Specifically, the acetone doped sample sintered at 700 °C has the best $J_c$ values in the magnetic field, even higher than that of the 5 at % C-added samples heated at 900°C presented in our previous work.[18] At 4.2 K, the $J_c$ of the acetone doped samples sintered at 700 °C reached up to $2.4 \times 10^4$ A/cm$^2$ at 10 T and $1 \times 10^4$ A/cm$^2$ at 12 T, respectively. At the same time, the acetone doped tapes showed a much smaller dependence of $J_c$ than the pure sample on magnetic field, which was almost the same as the C-doping sample. The result of the high $J_c$ values for the acetone doped samples suggests that the effective pinning centers were introduced in the superconducting cores.

Figure 2 shows the typical XRD patterns of the superconducting cores for MgB$_2$ samples doped with and without acetone. Clearly, MgB$_2$ is found to be the main phase in all samples, with minor impurity phases of MgO present. Mg peaks can be observed only in the sample sintered at 600 °C, as a result of the incomplete reaction. The dashed lines in figure 2 show the position of the (002) and (110) peaks of a pure MgB$_2$ sample. The position of the (002) peak shows almost no shift in acetone milled samples. Obviously, the position of (110) peak of all acetone doped samples shifts to higher angles, indicating a decrease of the *a*-axis of the crystal lattice. The shrinkage of *a* is attributed to the substitution of C for B.[5,19] We thought that the C element comes from the pyrolysis of acetone vapor, which is possibly in accordance with the following scheme above 500 °C:[20-22]

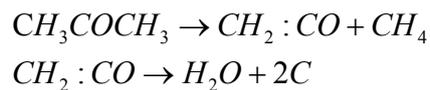
$$CH_3COCH_3 \rightarrow CH_2:CO + CH_4$$
$$CH_2:CO \rightarrow H_2O + 2C$$

The highly reactive C from the decomposition of acetone vapor can easily substitute into the lattice of MgB$_2$, which is in good agreement with the results of XRD. The inset of figure 2 shows the full width at half maximum (FWHM) value of the (110) peak of the different samples. It was found that the FWHM values of all the acetone doped samples were higher than that of pure one. This increase can imply a systematic reduction of $MgB_2$ grain size for acetone doped samples. Obviously, the FWHM values for acetone doped samples show a linear decrease with increasing sintering temperature, which suggests that the crystallinity of the $MgB_2$ core was improved at higher temperatures.[23]

Figure 3 shows the superconducting transition temperatures ($T_c$) for the acetone doped samples heat-treated from 600 to 800°C. The data of the undoped tape heat-treated at 800 °C is also included as a reference. As can be seen, the $T_c$ onset for the pure sample is 36.6 K, and the $T_c$ for doped tapes decrease to 32.08, 35.04, and 34.06 K with increasing the sintering temperature to 600, 700, and 800 °C, respectively. The drop of $T_c$ has also been observed in the bulk samples milled with acetone by other research group.[15] This lowering of the $T_c$ can be attributed to the higher oxygen mole fraction in doped samples and the substitution of C for B.[24] Meanwhile, a broadening of $\triangle T_c$=1.2 to 4.8, 2.23, and 2.5 K was observed, respectively. The increase of $\triangle T_c$ may point to an increasing inhomogeneity of the carbon distribution among the grains. The lowest $T_c$ value and the highest $\triangle T_c$ for the acetone doped sample sintered at 600 °C are probably the result of the incomplete reaction.

Microstructural analyses are employed to further elucidate the mechanism for the enhancement of $J_c$ in 5 wt % acetone doped samples. Figure 4(a) and 4(b) shows the typical SEM images of the fractured core layers for the samples with and without acetone dopant, respectively. The SEM results clearly reveal that the $MgB_2$ cores for all the tapes show a similar microstructure; however, the morphology of the acetone doped sample was refined to smaller grains compared to that of the pure tape, which is in good agreement with the results of the FWHM. The enhanced number of grain boundaries associated with the smaller grain size can improve the flux pinning ability, like in the case of high-$J_c$ $Nb_3Sn$ superconductors.[25] In the meantime, much larger melted regions of intergrains were observed in the acetone doped tapes, resulting in the better connectivity between the $MgB_2$ grains and enhanced $J_c$. On the other hand, the acetone doped samples had quite uniform microstructure with fewer voids, which also improved the linkages of grains. Figure 4(c) and 4(d) shows the typical low-magnification and high-resolution TEM micrographs for the acetone doped tapes sintered at 700 °C. As shown in low-magnification TEM micrographs, the samples are tightly packed $MgB_2$ nanoparticles with small grain size less than 50 nm, which agrees with the result of SEM. The high-resolution TEM image clearly shows a large number of dislocations within $MgB_2$ grains, which are similar to those shown in stearic acid and stearate doped samples.[11,12] Well defined ring patterns from the selected area electron diffraction patterns [inset of Fig. 4(d)] further demonstrate a very fine grain size in the doped tapes.

The change of the lattice parameter $a$ as a function of the additive acetone content is shown in Figure 5. As a reference, the $a$-axis length of the pure tape is also included. The obvious decrease of $a$-axis lengths for all the acetone doped tapes is an indication of carbon-for-boron substitution. The decrease of the $a$-axis length is almost independent of the amount of acetone, suggesting that more acetone addition did not proportionally lead to more carbon substitution and a saturated carbon substitution for boron has been reached in the 5 wt % acetone doped sample, which is similar to the results of other research groups.[17,26] The $T_c$ for the undoped and doped samples was determined by DC susceptibility measurements (Fig. 5). As can be seen, the $T_c$ decreases with the increase of additive content, which is similar to that of C doped samples.[6]

Figure 6 summarizes the $J_c$ at 4.2 K in magnetic fields up to 14 T for Fe-sheathed $MgB_2$ tapes with various amounts of acetone doping from 0 to 15 wt % that were heat treated at 700 °C. It is noted that the highest $J_c$ value of the Fe-sheathed tapes was achieved in the 5 wt % acetone addition, and then further increasing the additive content caused a reduction of $J_c$ in magnetic fields below 14 T. This decrease of $J_c$, especially in the low magnetic field region, suggests that the connectivity of $MgB_2$ grains were degraded by the further acetone doping, which is in good agreement with the results of $T_c$. On the other hand, MgO content between $MgB_2$ grains as insulating precipitate increases with increasing the additive content, thus it will reduce the effective cross-sectional area of the sample, and thereby block supercurrents. It should be noted that, the $J_c$ data for 5 wt % acetone doped sample in Fig. 6 measured at Grenoble High Magnetic Field Laboratory was almost the same as that measured at HFLSM in Sendai (Fig. 1), which indicates that our samples were well reproducible.

Clearly, the $J_c$ of $MgB_2$ tapes were strongly enhanced by acetone doping. This enhancement probably resulted from the increase of the flux pinning, which is attributed to two aspects: the substitute of C for B and the smaller $MgB_2$ grain size. The highly reactive C from the decomposition of acetone vapor can easily substitute into the lattice of $MgB_2$, which can be proved by the shrinkage of the *a*-axis for acetone doped samples. The smaller $MgB_2$ grain size could be observed by the SEM, TEM, and could also be proved by the higher FWHM values of all the acetone doped samples. Therefore, a high density of flux-pinning centers is responsible for the excellent performance in our acetone doped samples.

We could conclude that the acetone is both an effective ball-milling medium and liquid dopant: (1) The precursor powders can be mixed homogeneously by using acetone as a ball-milling medium. At the same time, we found that the wet pre-mixed powders are easier than the dry ones to fill into the iron tube. (2) The highly reactive C from the pyrolysis of acetone vapor can easily substitute into the lattice of $MgB_2$. Thus, the acetone is also an effective carbon source.

## 4. Conclusions

$MgB_2$/Fe tapes were fabricated with wet pre-mixed powders using acetone as ball-milling medium by the in-situ PIT process. It is found that all samples milled with acetone exhibited superior $J_c$ values compared to the pure tape. Specifically, the acetone milled sample sintered at 700 °C has the best $J_c$ values in the magnetic field, even higher than that of the C-doping sample sintered at 900 °C. Both the analysis results of XRD and the lowering of $T_c$ suggest that the highly reactive C from the pyrolysis of acetone substitute into the lattice of $MgB_2$. At the same time, the SEM and TEM results clearly reveal that smaller grains and a large number of dislocations within $MgB_2$ grains in the acetone doped tapes, result in the enhancement of the flux pining, thus the in-field $J_c$ for acetone doped samples was improved significantly.

**Acknowledgments**

The authors thank Haihu Wen, Liye Xiao, Ling Xiao, Liangzhen Lin, Honglin Du, Weiwen Huang, H. Matsuo, and T. Inoue for their help and useful discussion. This work is partially supported by the Beijing Municipal Science and Technology Commission under Grant No. Z07000300700703, National '973' Program (Grant No. 2006CB601004) and National '863' Project (Grant No. 2006AA03Z203).

# Captions

Figure 1    Transport critical current densities at 4.2 K as a function of magnetic field for $MgB_2$/Fe tapes milled with and without acetone. The measurements were performed in magnetic fields parallel to the tapes surface. The $I_c$ measurements were performed at HFLSM in Sendai.

Figure 2    XRD patterns of superconducting cores for $MgB_2$ samples milled with and without acetone. The data were obtained after peeling off the Fe-sheath.

Figure 3    Normalized magnetic susceptibility against temperature for the pure and 5 wt % acetone doped tapes. The inset shows an enlarged view near the superconducting transitions.

Figure 4    SEM and TEM images of the fractured $MgB_2$ core layers of Fe-sheathed tapes: (a) SEM micrograph for pure sample sintered at 800 °C. (b) SEM micrograph for 5 wt % acetone doped sample sintered at 700 °C. (c) Low-magnification TEM micrograph for 5 wt % acetone doped sample sintered at 700 °C. (d) High-magnification TEM micrograph for 5 wt % acetone doped sample sintered at 700 °C. The diffraction pattern of the sample is shown in the corner of the image (d).

Figure 5    Critical temperature onset and the $a$-parameter of the crystal lattice as function of the carbon content of $MgB_2$ samples with added acetone.

Figure 6    $J_c$–$B$ properties of different acetone doped level samples heated at 700 °C. The $I_c$ measurements were performed at Grenoble High Magnetic Field Laboratory.

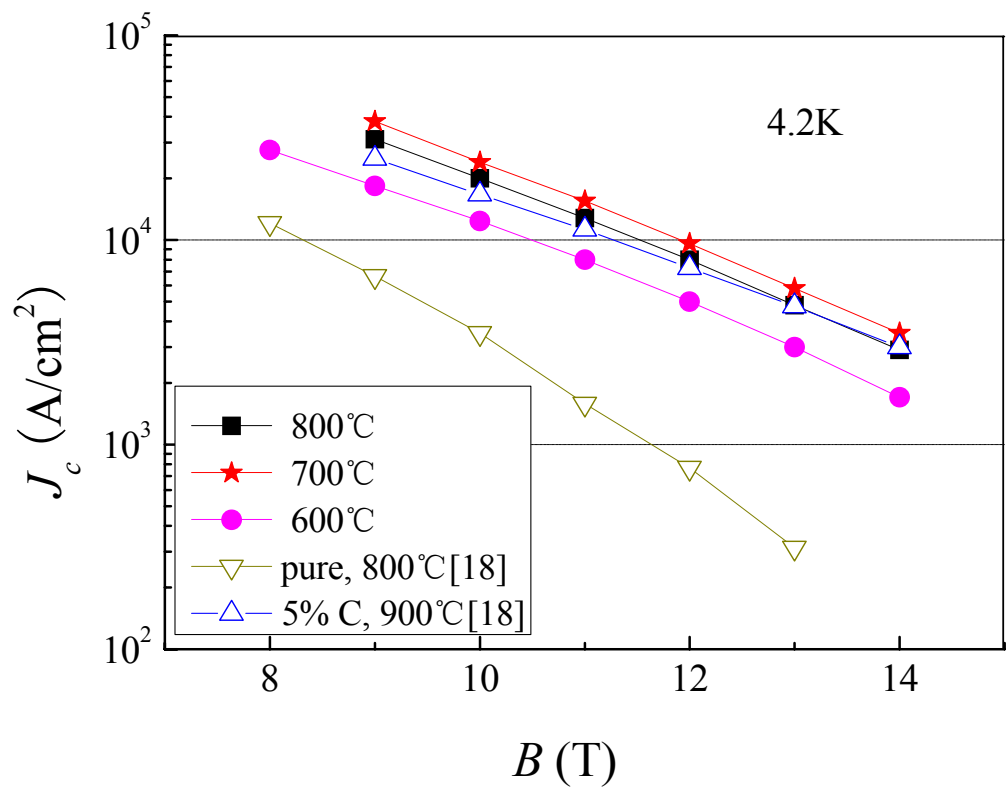

Fig.1 Wang et al.

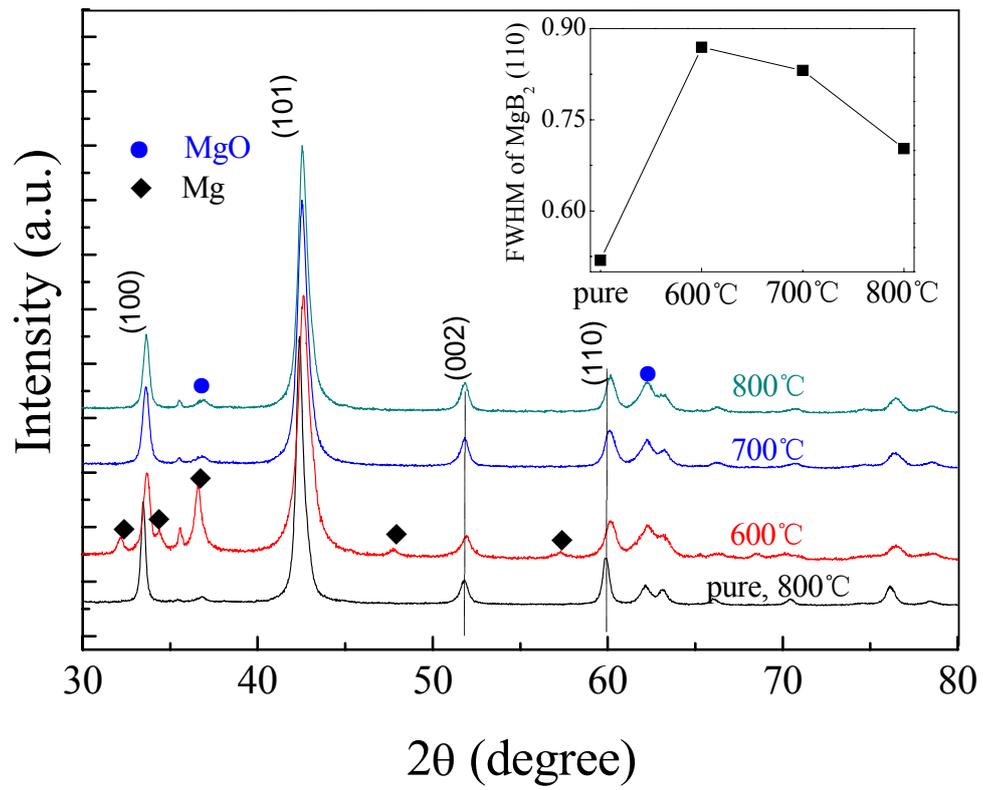

Fig.2 Wang et al.

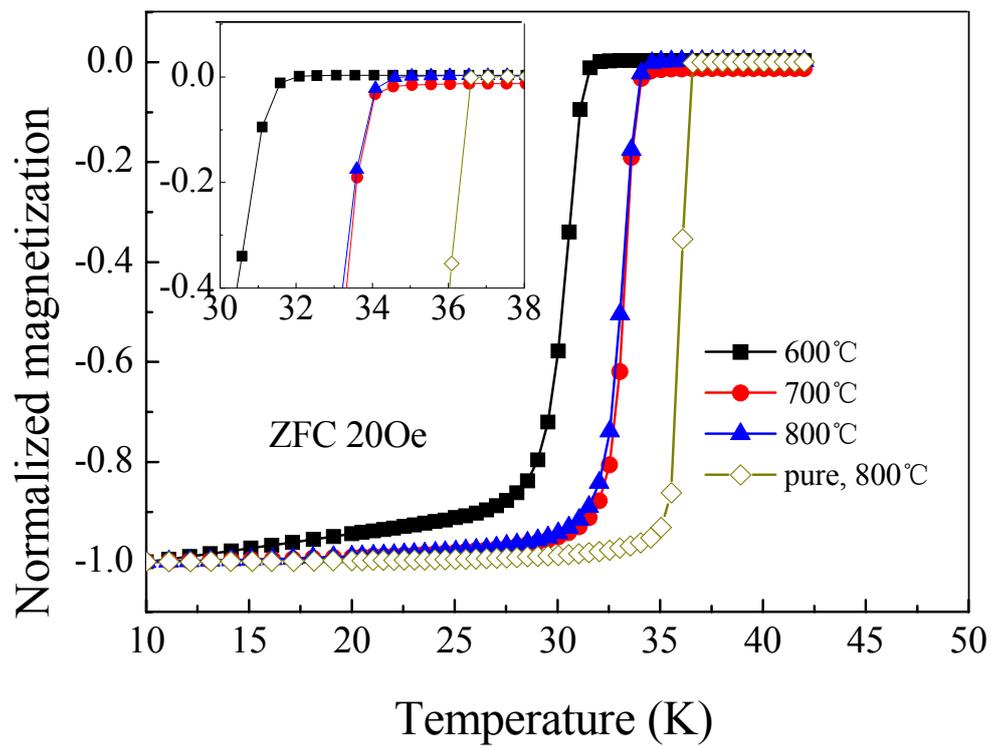

Fig.3 Wang et al.

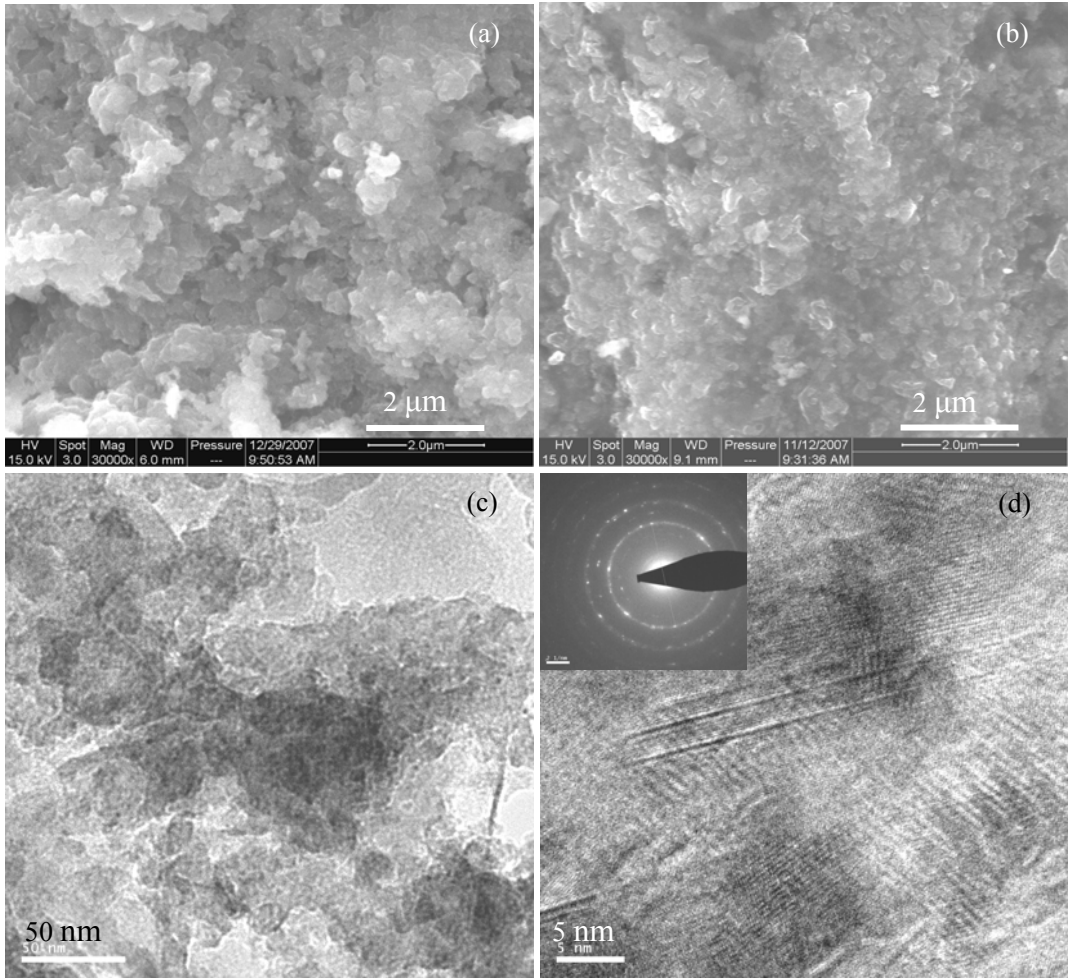

Fig.4 Wang et al.

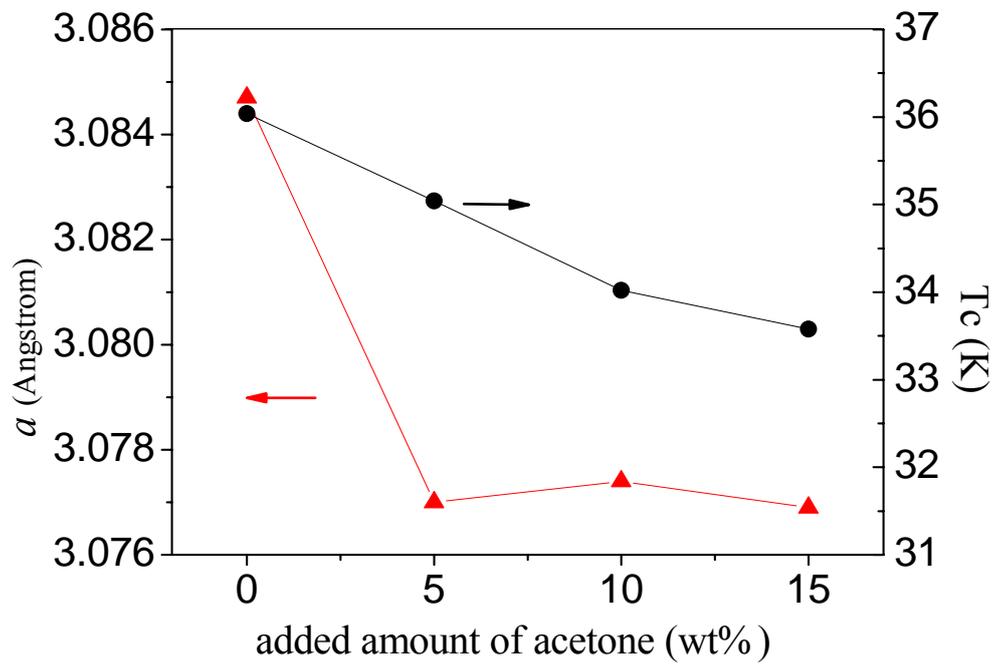

Fig.5 Wang et al.

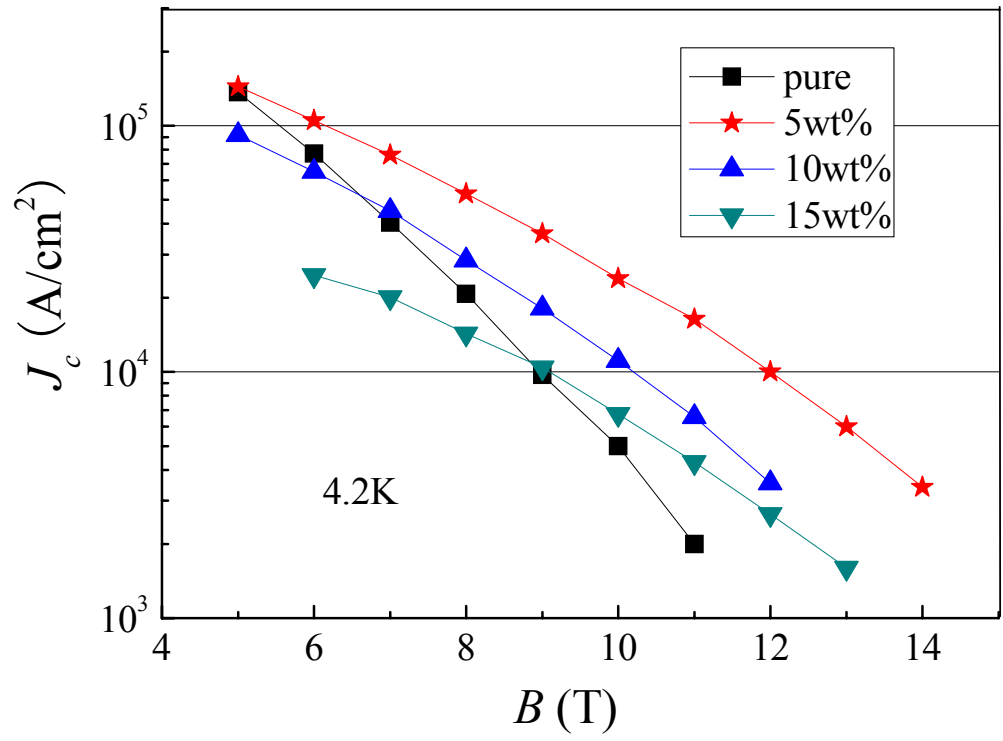

Fig.6 Wang et al.